%Paper: hep-th/9507153
%From: Bong Lian <lian@abel.math.harvard.edu>
%Date: Thu, 27 Jul 1995 10:12:17 -0400

\input harvmac

\def\frac#1#2{{#1\over#2}}

\def\journal#1&#2(#3){\unskip, #1~\bf #2 \rm(19#3) }
\def\andjournal#1&#2(#3){\sl #1~\bf #2 \rm (19#3) }

\catcode`\@=11\def\slash#1{\mathord{\mathpalette\c@ncel{#1}}}
\overfullrule=0pt
\def\steepslash{\c@ncel}
\def\frac#1#2{{#1\over #2}}

\def\:{\!:\!}
\def\inbar{\,\vrule height1.5ex width.4pt depth0pt}
\def\IQ{\relax\,\hbox{$\inbar\kern-.3em{\rm Q}$}}
\def\IB{\relax{\rm I\kern-.18em B}}
\def\IC{\relax\hbox{$\inbar\kern-.3em{\rm C}$}}
\def\IP{\relax{\rm I\kern-.18em P}}
\def\IR{\relax{\rm I\kern-.18em R}}
\def\ZZ{\relax\ifmmode\mathchoice
{\hbox{Z\kern-.4em Z}}{\hbox{Z\kern-.4em Z}}
{\lower.9pt\hbox{Z\kern-.4em Z}}
{\lower1.2pt\hbox{Z\kern-.4em Z}}\else{Z\kern-.4em Z}\fi}

\catcode`\@=12

%                      Zeitschriften:
\def\npb#1(#2)#3{{ Nucl. Phys. }{B#1} (#2) #3}
\def\plb#1(#2)#3{{ Phys. Lett. }{#1B} (#2) #3}
\def\pla#1(#2)#3{{ Phys. Lett. }{#1A} (#2) #3}
\def\prl#1(#2)#3{{ Phys. Rev. Lett. }{#1} (#2) #3}
\def\mpla#1(#2)#3{{ Mod. Phys. Lett. }{A#1} (#2) #3}
\def\ijmpa#1(#2)#3{{ Int. J. Mod. Phys. }{A#1} (#2) #3}
\def\cmp#1(#2)#3{{ Comm. Math. Phys. }{#1} (#2) #3}
\def\cqg#1(#2)#3{{ Class. Quantum Grav. }{#1} (#2) #3}
\def\jmp#1(#2)#3{{ J. Math. Phys. }{#1} (#2) #3}
\def\anp#1(#2)#3{{ Ann. Phys. }{#1} (#2) #3}
\def\prd#1(#2)#3{{ Phys. Rev. } {D{#1}} (#2) #3}
\def\ptp#1(#2)#3{{ Progr. Theor. Phys. }{#1} (#2) #3}
\def\aom#1(#2)#3{{ Ann. Math. }{#1} (#2) #3}

\def\C{{\bf C}}

\def\P{{\bf P}}

\def\cL{{\cal L}}

\def\cicy#1(#2|#3)#4{\left(\matrix{#2}\right|\!\!
                     \left|\matrix{#3}\right)^{{#4}}_{#1}}

\def\qed{{\bf $\bullet$}}

\global\newcount\thmno \global\thmno=0
\def\question#1{\global\advance\thmno by1
\bigskip\noindent{\bf Question \secsym\the\thmno. }{\it #1}
\par\nobreak\medskip\nobreak}
\def\theorem#1{\global\advance\thmno by1
\bigskip\noindent{\bf Theorem \secsym\the\thmno. }{\it #1}
\par\nobreak\medskip\nobreak}
\def\proposition#1{\global\advance\thmno by1
\bigskip\noindent{\bf Proposition \secsym\the\thmno. }{\it #1}
\par\nobreak\medskip\nobreak}
\def\corollary#1{\global\advance\thmno by1
\bigskip\noindent{\bf Corollary \secsym\the\thmno. }{\it #1}
\par\nobreak\medskip\nobreak}
\def\lemma#1{\global\advance\thmno by1
\bigskip\noindent{\bf Lemma \secsym\the\thmno. }{\it #1}
\par\nobreak\medskip\nobreak}
\def\conjecture#1{\global\advance\thmno by1
\bigskip\noindent{\bf Conjecture \secsym\the\thmno. }{\it #1}
\par\nobreak\medskip\nobreak}
\def\exercise#1{\global\advance\thmno by1
\bigskip\noindent{\bf Exercise \secsym\the\thmno. }{\it #1}
\par\nobreak\medskip\nobreak}
\def\remark#1{\global\advance\thmno by1
\bigskip\noindent{\bf Remark \secsym\the\thmno. }{\it #1}
\par\nobreak\medskip\nobreak}
\def\problem#1{\global\advance\thmno by1
\bigskip\noindent{\bf Problem \secsym\the\thmno. }{\it #1}
\par\nobreak\medskip\nobreak}
\def\proof{\noindent Proof: }

\def\thmlab#1{\xdef
#1{\secsym\the\thmno}\writedef{#1\leftbracket#1}\wrlabeL{#1=#1}}
%
% redefine \newsec so that all \thmno set to zero in a new section
%
%\def\newsec#1{\global\advance\secno by1\message{(\the\secno. #1)}
%%\ifx\answ\bigans \vfill\eject \else \bigbreak\bigskip \fi  %if desired
%\global\subsecno=0\thmno=0\eqnres@t\noindent{\bf\the\secno. #1}
%\writetoca{{\secsym} {#1}}\par\nobreak\medskip\nobreak}
\def\eqnres@t{\thmno=0%
\xdef\secsym{\the\secno.}\global\meqno=1\bigbreak\bigskip}
\def\sequentialequations{\def\eqnres@t{\bigbreak}}\xdef\secsym{}
\def\npb#1(#2)#3{{ Nucl. Phys. }{B#1} (#2) #3}
\def\plb#1(#2)#3{{ Phys. Lett. }{#1B} (#2) #3}
\def\pla#1(#2)#3{{ Phys. Lett. }{#1A} (#2) #3}
\def\prl#1(#2)#3{{ Phys. Rev. Lett. }{#1} (#2) #3}
\def\mpla#1(#2)#3{{ Mod. Phys. Lett. }{A#1} (#2) #3}
\def\ijmpa#1(#2)#3{{ Int. J. Mod. Phys. }{A#1} (#2) #3}
\def\cmp#1(#2)#3{{ Commun. Math. Phys. }{#1} (#2) #3}
\def\cqg#1(#2)#3{{ Class. Quantum Grav. }{#1} (#2) #3}
\def\jmp#1(#2)#3{{ J. Math. Phys. }{#1} (#2) #3}
\def\anp#1(#2)#3{{ Ann. Phys. }{#1} (#2) #3}
\def\prd#1(#2)#3{{ Phys. Rev.} {D\bf{#1}} (#2) #3}

\def\Tx{\Theta_x}
\def\Ty{\Theta_y}

\def\C{{\bf C}}

\def\P{{\bf P}}

\def\cL{{\cal L}}
\def\cR{{\cal R}}

\def\inbar{\,\vrule height1.5ex width.4pt depth0pt}
\def\IQ{\relax\,\hbox{$\inbar\kern-.3em{\rm Q}$}}
\def\IB{\relax{\rm I\kern-.18em B}}
\def\IC{\relax\hbox{$\inbar\kern-.3em{\rm C}$}}
\def\IP{\relax{\rm I\kern-.18em P}}
\def\IR{\relax{\rm I\kern-.18em R}}
\def\ZZ{\relax\ifmmode\mathchoice
{\hbox{Z\kern-.4em Z}}{\hbox{Z\kern-.4em Z}}
{\lower.9pt\hbox{Z\kern-.4em Z}}
{\lower1.2pt\hbox{Z\kern-.4em Z}}\else{Z\kern-.4em Z}\fi}

\lref\cdgp{P. Candelas, X. De la Ossa, P. Green and L. Parkes, Nucl. Phys. B359
(1991) 21.}\lref\vk{C. Vafa and S. Kachru, {\sl Exact Results
for N=2 Compactifications of Heterotic Strings}, hep-th/9505105.}
\lref\dWKLL{B.\ de Wit, V.\ Kaplunovsky, J.\ Louis and D.\
L\"{u}st, {\it Perturbative Couplings of Vector Multiplets in $N=2$
Heterotic String Vacua}, hep-th/9504006.}
\lref\AFGNT{I.\ Antoniadis, S.\ Ferrara,
E.\ Gava, K.S.\ Narain and T.R.\ Taylor, {\it Perturbative
Prepotential and Monodromies in N=2 Heterotic Superstring},
hep-th/9504034.}
\lref\HKTYI{S.\ Hosono, A.\ Klemm, S.\ Theisen and S.-T. Yau,
Comm. Math. Phys. 167(1995) 301, hep-th/9308122.}
\lref\COFKM{P.\ Candelas, X.\ de la Ossa, A.\ Font, S.\ Katz and D.\
Morrison, Nucl. Phys. 416 (1994) 481, hep-th/9308083.}
\lref\lkm{A. Klemm, W. Lerche and P. Mayr, {\sl K3 fibrations and
heterotic type II string duality}, hep-th/9506112.}
\lref\ly{B. Lian and S.-T. Yau, {\sl Arithmetic Properties
of Mirror Maps and Quantum Couplings}, hep-th/9411234.}
\lref\lkt{V. Kaplunovsky, J. Louis and S. Theisen, {\sl Aspects of duality in
N=2 string vacua}, hep-th/9506110.}
\lref\CN{J.H. Conway and S.P. Norton,``Monstrous Moonshine'', Bull. London
Math. Soc., 11(1979) 308-339.}
\lref\gp{B. Greene and R. Plesser, Nucl. Phys. B 338 (1990),15-37.}
\lref\emm{Essays on Mirror Manifolds, Ed. S.-T. Yau,
(International Press, Hong Kong 1992)}
\lref\lianyaufour{B.H. Lian and S.-T. Yau, {\sl
Mirror Maps, Modular Relations and Hypergeometric Series I}, hep-th/9507151.}
\lref\yonemura{T. Yonemura, Tohoku Math. J. 42 (1990) 351-380.}
\lref\lkt{V. Kaplunovsky, J. Louis and S. Theisen, {\sl Aspects of duality in
N=2 string vacua}, hep-th/9506110.}
\lref\vafawitten{C. Vafa and E. Witten, {\sl Dual String Pairs with
$N=1$ and $N=2$ Supersymmetry in Four Dimensions}, hep-th/9507050.}

% when printing final draft, remove \draftmode
%\draftmode
%q-alg/9507153
\Title{}{Mirror Maps, Modular Relations and Hypergeometric Series II}

\centerline{
Bong H. Lian$^{1}$\footnote{}{$^1$~~Department of Mathematics,
Brandeis University, Waltham, MA 02154.}
 and Shing-Tung Yau$^2$\footnote{}{$^2$~~Department of Mathematics,
Harvard University, Cambridge, MA 02138.} }
\vskip .2in

Abstract. As a continuation of \lianyaufour, we
study modular properties of the periods, the mirror maps and Yukawa couplings
for multi-moduli Calabi-Yau varieties.
In Part A of this paper, motivated by the recent work
of Kachru-Vafa, we degenerate a three-moduli
family of Calabi-Yau toric varieties along a codimension one
 subfamily which can be described
by the vanishing of certain Mori coordinate, corresponding to going to
the ``large volume limit'' in a certain direction.
Then we see that the deformation space of the subfamily is the same as a
certain
family of K3 toric surfaces. This family can
in turn be studied by further degeneration along a subfamily which in the end
is described by a family of elliptic curves. The
periods of the K3 family (and hence the original Calabi-Yau family) can be
described by the squares of the periods of the elliptic curves.
The consequences include: (1) proofs of various
conjectural formulas of physicists \vk\lkm~
involving mirror maps and modular functions;
(2) new identities involving multi-variable hypergeometric series and
modular functions -- generalizing \lianyaufour.
In Part B, we study for two-moduli families the perturbation series of
the mirror map and the type A Yukawa couplings near certain large
volume limits. Our main tool is a new class of polynomial
PDEs associated with Fuchsian PDE systems. We derive the first
few terms in the perturbation series.
For the case of degree 12 hypersurfaces in $\P^4[6,2,2,1,1]$,
in one limit the series of the couplings are expressed
in terms of the $j$ function. In another limit, they are expressed
in terms of rational functions. The latter give
explicit formulas for infinite sequences of ``instanton numbers'' $n_d$.

\Date{hep-th/9507153} %replace this line by \draft  for preliminary versions
             %or specify \draftmode at some point

%if you want double-space, use e.g. \baselineskip=20pt plus 2pt minus 2pt

\vfill\eject

This paper is a continuation of \lianyaufour. There we study the modular
properties of multi-moduli families Calabi-Yau varieties degenerated along a
dimension one subfamilies. In this article, we study properties of the mirror
map, periods and the type A Yukawa couplings under a degeneration along
a codimension one subfamily, and their perturbations around
this subfamily. The problem is clearly motivated by
recent developments in the so-called heterotic-type II string duality.

\newsec{Part A: Degree 24 hypersurfaces in $\P^4[1,1,2,8,12]$}

The mirror symmetry of this
family of Calabi-Yau toric varieties $X$ has been studied in detail in
\HKTYI. Its Picard-Fuchs system is given by ($\Theta_x=x{d\over dx}$ etc.)
\eqn\dumb{\eqalign{
L_1=&~\Theta_x(\Theta_x-2\Theta_z)-12x(6\Theta_x+5)(6\Theta_x+1)\cr
L_2=&\Theta_y^2-y(2\Theta_y-\Theta_z+1)(2\Theta_y-\Theta_z)\cr
L_3=&\Theta_z(\Theta_z-2\Theta_y)-z(\Theta_z-\Theta_x+1)(2\Theta_z-\Theta_x).
}}
The $x,y,z$ are deformation coordinates, which we call
the Mori coordinates (see \HKTYI~ for definition),
near the ``large volume limit'' in the family of Calabi-Yau varieties.

Comparing the type II string compactification along $X$
 with a heterotic string theory,
Kachru-Vafa suggest that one should study the limit $y\rightarrow0$.
When restricted to this subfamily with $y=0$,
a subset of the periods of this subfamily satisfy a new system
given by
\eqn\kthreeI{\eqalign{
L_1=&~\Theta_x(\Theta_x-2\Theta_z)-12x(6\Theta_x+5)(6\Theta_x+1)\cr
L_3=&\Theta_z^2-z(2\Theta_z-\Theta_x+1)(2\Theta_z-\Theta_x). }}
This is identical to the Picard-Fuchs system for the
family of toric K3 surfaces corresponding to
degree 12 hypersurfaces in $\P^3[1,1,4,6]$. (For the relevance of
the appearance of K3 surfaces and their moduli spaces in heterotic-type II
duality, see the recent papers \lkm\vafawitten.)
By further restricting along $z=0$, the Picard-Fuchs system reduces to
a single equation
\eqn\ellipticI{L=\Theta_x^2-12x(6\Theta_x+5)(6\Theta_x+1).}
This is the Picard-Fuchs operator for a family of elliptic curves
in $\P^2[1,2,3]$. This suggests a close relationship between the curves
and the above 2-moduli K3 family, and ultimately the 3-moduli family
of Calabi-Yau varieties $X$. Indeed, it is found numerically in \lkm~
that the mirror map defined by the K3 family can be given in terms
of the $j$-function.

In this section, we will prove that the solutions to \kthreeI
are given by  "squares'' of solutions to \ellipticI. This result
(1) generalizes a theorem in \lianyaufour~ which we will review briefly;
(2) proves some formulas of \lkm~ as a consequence.

Consider the differential operators
\eqn\dumb{\eqalign{
\ell=&~\Theta_x^3-\sum_{i=1}^m \lambda_ix^i(\Theta_x+i/2)
(\Theta_x+i/2+\nu_i)(\Theta_x+i/2-\nu_i)\cr
\tilde{\ell}=&~\Theta_x^2-\sum_{i=1}^m \lambda_ix^i
(\Theta_x+i/4+\nu_i/2)(\Theta_x+i/4-\nu_i/2) }}
where the $\lambda_i,\nu_i$ are arbitrary complex numbers.
In \lianyaufour, we prove that if $\tilde{\ell}f(x)=0$ then
$\ell f(x)^2=0$. This result was inspired by the observation in
numerous examples (see \lianyaufour) that the periods of certain
1-moduli K3 family are nothing but products of periods of some
family of elliptic curves. This leads to nontrivial identities
involving modular functions and series solutions to an ODE of
Fuchsian type. Explicitly if $w_0,w_1$ form
a basis of solutions to $\tilde\ell$, then we get a basis of solutions
to $\ell$: $w_0^2, w_0w_1, w_1^2$.
 This suggests to us a
2-moduli analogue for the systems \kthreeI, \ellipticI: if
$w_0(x),w_1(x)$ are solutions to \ellipticI, are the $w_i(x)w_j(z)$
solutions to \kthreeI ? The answer turns out to be no, but almost.
Note that the span of $w_i(x)w_j(z)$ is 4 dimension. This is the same
as the dimension of the solution space for \kthreeI
with at most $Log$ singularities.

\theorem{Let $L_1,L_3,L$ be as defined in \kthreeI, \ellipticI.
There exists a rational mapping $\C^2\rightarrow\C^2$, $(R,S)\mapsto(x,z)$,
such that if $f(x)$ is a solution to $L$,
then $f(R(x,z))f(S(x,z))$ is a solution to $L_1,L_3$,
where $(x,z)\mapsto (R(x,z),S(x,z))$ is (any branch of) the inverse mapping.}

\proof We will construct the mapping using the condition that
\eqn\Lff{L_1f(R(x,z))f(S(x,z))\equiv0~~mod~ Lf(R),Lf(S).}
It will be seen that the mapping we will construct also
satisfies the analogous condition for $L_2$.

Clearly by expanding the expression $L_1f(R(x,z))f(S(x,z))$
by chain rule, we get
a homogeneuos quadratic polynomial of
$f^{(i)}(R), f^{(i)}(S)$, whose coefficients
are differential polynomials of $R(x,z), S(x,z)$. Upon applying the
conditions that $Lf(R)=0=Lf(S)$, we can eliminate any appearance
of $f''(R),f''(S)$. Thus after the elimination,
a sufficient condition for \Lff to hold
is that coefficients of $f(R)f(S), f'(R)f(S), f(R)f'(S), f'(R)f'(S)$,
each vanishes identically. Thus we want to solve the
conditions of the vanishing of these
coefficients, and they are given by:
\eqn\SRPDE{\eqalign{
(1)~~ &-R S x + 1728 R^2  S x + 1728 R S^2  x - 2985984 R^2  S^2  x - \cr
& 2 S R_z R_x + 3456 S^2  R_z R_x +
S R_x^2  - 1728 S^2  R_x^2  - 432 S x R_z^2  + \cr
  &    746496 S^2  x R_x^2  - 2 R S_z S_x+
        3456 R^2  S_z   S_x + R S_x^2  - \cr
&       1728 R^2  S_x^2   - 432 R x S_x^2   +
     746496 R^2  x S_x^2 =0\cr
(2)~~ & 432 R x R_x - 746496 R^2  x R_x -
  2 R_z R_x + 5184 R R_z R_x +
  R_x^2  - \cr & 2592 R R_x^2  - 432 x R_x^2  +
   1119744 R x R_x^2  + 2 R R_{xz} - 3456 R^2  R_{xz}   - \cr
 &     R R_{xx}+ 1728 R^2  R_{xx} + 432 R x R_{xx} -
    746496 R^2  x R_{xx}   =0\cr
(3)~~ & {\sl As~in~(2)~with~S,R~interchanged.}\cr
(4)~~ &  -S_z R_x  - R_z  S_x +       R_x S_x - 432 x R_x S_x=0 }}
where $S_x$ means $\Theta_xS$ etc.
This is an overdetermined system of polynomial PDEs.
We claim that the following relations define an algebraic solution
to \SRPDE:
\eqn\RSxz{\eqalign{
&~R+S-864RS-x=0\cr
&RS(1-432R)(1-432S)-x^2z=0. }}
The proof is by direct computation.
(This solution is easy to motivated by the following consideration.
Since we propose that the periods of the K3 surfaces in question
 are symmetric squares of those of the elliptic curves, it is reasoable that
the K3 moduli $x,z$ are symmetric functions in the elliptic curve moduli $R,S$.
The above solution makes $x,z$ the simplest kinds of algebraic symmetric
functions of $R,S$.)
Note that \RSxz~ defines a rational
mapping $(R,S)\mapsto(x,z)$. It is also easy to check that given
this solution, the condition
\eqn\Lff{L_2f(R(x,z))f(S(x,z))\equiv0~~mod~ Lf(R),Lf(S)}
holds automatically. This completes our proof. \qed

Note that by linearity, if $f,g$ are any solutions to \ellipticI,
$f(R(x,z))g(S(x,z))$ is a solution to \kthreeI.
As a first consequence, we prove a formula first conjectured to exist
in \vk on physical ground, and then found numerically in \lkm.

\corollary{The mirror map for \kthreeI~ is given by
\eqn\xzj{\eqalign{
x(q_1,q_3)=&~ 2{1/j(q_1)+1/j(q_1q_3)-1728/(j(q_1)j(q_1q_3))\over
1+\sqrt{1-1728/j(q_1)}\sqrt{1-1728/j(q_1q_3)} }\cr
z(q_1,q_3)=&~{1\over j(q_1)j(q_1q_3) x(q_1,q_3)^2}. }}
}

\proof Recall that \kthreeI has a {\it unique} powers series solution $w_0$
with leading term $1$, {\it unique} solutions $w_1, w_3$ of the form
$w_1=Log~x+g_1, w_3=Log~z+g_3$ with $g_1, g_3\rightarrow0$
as $|x|,|z|\rightarrow0$.
(This is so because this is a holonomic PDE system with regular
singularity at $x=0,z=0$. It is also straightforward to check this
directly using \kthreeI~ and the resulting recursion relations on
the coefficients of the power series.)
The mirror map $x(q_1,q_3), z(q_1,q_3)$
for \kthreeI is then defined by the inverse of the power
series relations:
\eqn\dumb{\eqalign{q_1=&~e^{w_1/w_0}\cr
q_3=&~e^{w_3/w_0}. }}

Similarly, the ODE \ellipticI has a power series solution $\tilde{w}_0$
with leading term $1$ and a solution $\tilde{w}_1$ of the form
$Log~x+\tilde{g}$ with $g\rightarrow0$ as $|x|\rightarrow0$.
The mirror map $r(q)$ for \ellipticI is then defined by the inverse of the
power series relation $q=e^{\tilde{w}_3(r)/\tilde{w}_0(r)}$.
It is also easy to prove that (see \ly) $j(q)={1\over r(q)(1-432r(q))}$.

By the theorem above,
the following are three solutions to \kthreeI:
$\tilde{w}_0(R(x,z))\tilde{w}_0(S(x,z))$,
$\tilde{w}_1(R(x,z))\tilde{w}_0(S(x,z))$,
$\tilde{w}_0(R(x,z))\tilde{w}_1(S(x,z))-
\tilde{w}_1(R(x,z))\tilde{w}_0(S(x,z))$.
It is straightforward to solve
 \RSxz~for $R,S$ as power series in $x,z$, and we see there
are four branches of solutions. One branch has leading terms
$R=x+O(h^2), S=xz+O(h^3)$. (Here $O(h^k)$ means terms of total degree $k$
or higher.) The second branch has $R=1/432 +O(h), S=1/432+O(h)$.
The third and the fourth branches are obtained by interchanging the
roles of $R,S$ in the first two branches. We choose $R(x,z), S(x,z)$
to be given by the first branch. Then it easy to see that the three
solutions $\tilde{w}_0(R(x,z))\tilde{w}_0(S(x,z))$,
$\tilde{w}_1(R(x,z))\tilde{w}_0(S(x,z))$,
$\tilde{w}_0(R(x,z))\tilde{w}_1(S(x,z))-
\tilde{w}_1(R(x,z))\tilde{w}_0(S(x,z))$, have the leading
behaviour identical to that of the solutions $w_0,w_1,w_3$ respectively.
By uniqueness, we conclude that
\eqn\www{\eqalign{
w_0=&~\tilde{w}_0(R(x,z))\tilde{w}_0(S(x,z))\cr
w_1=&~\tilde{w}_1(R(x,z))\tilde{w}_0(S(x,z))\cr
w_3=&~\tilde{w}_0(R(x,z))\tilde{w}_1(S(x,z))-
\tilde{w}_1(R(x,z))\tilde{w}_0(S(x,z)). }}
This implies that
\eqn\dumb{\eqalign{
q_1=&~e^{\tilde{w}_1(R(x,z))/\tilde{w}_0(R(x,z))}\cr
q_1q_3=&~e^{\tilde{w}_0(S(x,z))/\tilde{w}_1(S(x,z))}. }}
Inverting these, applying \RSxz, and using $j(q)={1\over r(q)(1-432r(q))}$,
we see that \xzj~ follows. \qed

\corollary{Let $E_4$ be the normalized Eisenstein series of weight 4. Then
\eqn\cortwo{\eqalign{
&\left(\sum_{k=0}^\infty\sum_{m=0}^\infty\right. \left(\matrix{6k+12m \cr
3k+6m}\right)
\left(\matrix{3k+6m \cr 2k+4m}\right) \left(\matrix{k+2m \cr 2m}\right)
\left(\matrix{2m \cr m}\right) {1\over j(q_1)^m j(q_2)^m}\cr
 &\left. ~~~\times
\left(2{j(q_1)+j(q_2)-1728\over j(q_1)j(q_2)
+\sqrt{j(q_1)(j(q_1)-1728)}\sqrt{j(q_2)(j(q_2)-1728)} }\right)^k
 \right)^4
=E_4(q_1)E_4(q_2).} }
}

\proof Computing the power series solution $w_0$ to \kthreeI~
with leading term 1, we get
\eqn\dumb{w_0(x,z)=\sum_{n\geq 2m\geq0}
{(6n)!\over(3n)!(2n)!(m!)^2(n-2m)!}x^nz^m.}
Now  do a change of variable on the summation $n=k+2m$, put $q_2=q_1q_3$,
and apply the corollary above. We see that the
left hand side of \cortwo~ is $w_0(x(q_1,q_3),z(q_1,q_3))$.

By the first equation in \www, it is enough to show that
$\tilde{w}_0(r(q))^4=E_4(q)$ where
$\tilde{w}_0(x)$ is the solution to \ellipticI~
regular at $x=0$. In \lianyaufour, we have proved that
\eqn\dumb{\tilde{w}_0(r(q))^2={\Theta_q r(q)\over r(q)(1-432r(q))}.}
But we know that
\eqn\dumb{\eqalign{
j(q)=&~{1\over r(q)(1-432r(q))}\cr
E_4(q)=&~{(\Theta_q j(q))^2\over j(q)(j(q)-1728)^2}. }}
Combining the three equations, we get $E_4(q)=\tilde{w}_0(r(q))^4$. \qed

\newsec{Generalizations}

The technique we have used to study the example above is clearly applicable
to a more general class of PDEs. The only step which involves details
of the example is the system \SRPDE. It turns out that even the
form of our solution
\RSxz~ to \SRPDE~ has more general applicability as we now show.

Consider the PDE system
\eqn\generalPDE{\eqalign{
L_1=&~\Theta_x(\Theta_x-2\Theta_z)-
\lambda x(\Theta_x+1/2+\nu)(\Theta_x+1/2-\nu)\cr
L_3=&\Theta_z^2-z(2\Theta_z-\Theta_x+1)(2\Theta_z-\Theta_x) }}
and the ODE
\eqn\generalODE{L=\Theta_x^2-\lambda x(\Theta_x+1/2+\nu)(\Theta_x+1/2-\nu)}
where $\lambda,\nu$ are complex numbers. For $(\lambda,\nu)=(432,1/3)$
we recover the case \kthreeI~ above.
We now have the following generalization.

\theorem{The rational mapping $\C^2\rightarrow\C^2$, $(R,S)\mapsto(x,z)$,
defined by the relations
\eqn\dumb{\eqalign{
&~R+S-2\lambda RS-x=0\cr
&RS(1-\lambda R)(1-\lambda S)-x^2z=0 }}
has the following property:
if $f(x)$ is a solution to $L$,
then $f(R(x,z))f(S(x,z))$ is a solution to $L_1,L_3$,
where $(x,z)\mapsto (R(x,z),S(x,z))$ is (any branch of) the inverse mapping.}

The proof is vitually word for word similar to the proof
of the special case above.

Consider the following two families of toric K3 surfaces corresponding to
degree 6 hypersurfaces in $\P^3[1,1,2,2]$ and degree 8 hypersurfaces
in $\P^3[1,1,2,4]$ respectively. Their Picard-Fuchs systems are exactly
\generalPDE~ with $(\lambda,\nu)=(27,1/6), (64,1/4)$ respectively.
The corresponding ODEs are \generalODE with those parameter values.
It turns out that they
are exactly the Picard-Fuchs equations for two families
 of elliptic curves: degree 3 curves in $\P^2[1,1,1]$ and
degree 4 curves in $\P^2[1,1,2]$ respectively.
As shown in \ly, The mirror maps for these two families of curves are
hauptmoduls for the genus zero groups $\Gamma_0(2), \Gamma_0(3)$ respectively.

Finally it is amusing to note that the three examples above
with parameter values $(\lambda,\nu)= (27,1/6), (64,1/4), (432,1/3)$
correspond
to the so-called simple elliptic singularities
of types $E_6, E_7, E_8$ respectively.
(See introduction of \yonemura.) That is, the three families of
elliptic curves mentioned above -- degrees 3, 4, 6
hypersurfaces in $\P^2[1,1,1], \P^2[1,1,2], \P^2[1,2,3]$ respectively --
correspond to singularities of these types.
Note that their two dimensional counterparts are the three families of
K3 surfaces above -- degrees 6, 8, 12 hypersurfaces in $\P^3[1,1,2,2],
\P^3[1,1,2,4], \P^3[1,1,4,6]$ respectively.

It turns out that there is an explicit relation
between a generic member of the K3 family in $\P^3[1,1,2,2]$,
and a cubic family in $\P^2[1,1,1]$. That is, the surface
\eqn\dumb{y_1^6+y_2^6+y_3^3+y_4^3+a y_1y_2y_3y_4+by_1^3y_2^3=0}
(away from the hyperplane $y_1=0$) is a two-fold cover
of the cubic family
\eqn\dumb{(1+\lambda^6+b\lambda^3)x_1^3+x_2^3+x_3^3+a\lambda x_1x_2x_3=0}
via the map $(y)\mapsto(\lambda=y_2/y_1; x_1=y_1^2, x_2=y_3, x_3=y_4)$.
There is an analogous relation between the K3 family in
$\P^3[1,1,2,4]$ and a quartic family in $\P^2[1,1,2]$,
and similarly for $\P^3[1,1,4,6]$ and $\P^2[1,2,3]$.

There are also three dimensional Calabi-Yau varieties which bear
the same relation to the above K3 families as these K3 surfaces
bear with their elliptic curve counterparts. Namely the three K3 families
above correspond respectively to the
degrees 12, 16, 24 hypersurfaces in $\P^4[1,1,2,4,4],
\P^4[1,1,2,4,8], \P^4[1,1,2,8,12]$ respectively. A generic member of
the Calabi-Yau family of the form
\eqn\dumb{
z_1^{12}+z_2^{12}+z_3^6+z_4^3+z_5^3+az_1z_2z_3z_4z_5+bz_1^6z_2^6=0}
 in $\P^4[1,1,2,4,4]$ is a two-fold cover
of a sextic family in $\P^3[1,1,2,2]$, and similarly for the other
two cases.

\newsec{Part B: Schwarzian equations for linear PDEs}

Since the periods, the mirror map and the Yukawa couplings
are of fundamental importance for
the prediction of the numbers of rational curves via mirror symmetry,
one must understand these objects from as many points of views as one can.
For example, can one give an analytic characterization for the mirror map?
It has been shown in 1-modulus cases that the answer is yes: the mirror map
is characterized by some polynomial ODE near the ``large volume limit'' \ly.
Motivated by this problem, we study in this section the analogues in 2-moduli
cases. We will construct polynomial PDE systems naturally associated with
the Picard-Fuchs systems
for the periods of Calabi-Yau varieties. Later we will see that these PDEs
give us powerful tool for doing perturbation theory on the
periods, the mirror map and the type A Yukawa couplings.

Consider for fixed $m\geq2$ the following pair of linear partial differential
operators:
\eqn\generaltwo{\eqalign{
L_1=&~\sum_{0\leq i+j\leq m} a_{ij}\partial_x^i\partial_y^j\cr
L_2=&~\sum_{0\leq i+j\leq2} b_{ij}\partial_x^i\partial_y^j }}
where the $a_{ij}, b_{ij}$ rational function of $x,y$.
We assume that $b_{11}^2-4b_{02}b_{20}$ is not identically zero,
and that near $x=0=y$
the system admits unique series solutions $w_0, w_1,w_2$
with the leading behaviour
\eqn\dumb{\eqalign{
w_0(x,y)=&~1+O(h)\cr
w_1(x,y)=&~w_0 Log~x +O(h)\cr
w_2(x,y)=&~w_0 Log~y +O(h)}}
where $O(h^k)$ here means terms in powers of $x,y$ which are of total degree
at least $k$. We remark that all known examples of families of
Calabi-Yau varieties result in Picard-Fuchs systems of this kind, where
$x,y$ denotes the Mori coordinates for the complex structure deformation
space near the large volume limit.
Let $s=w_1/w_0$, $t=w_2/w_0$ as (locally defined) $\C^2$-valued maps of $x,y$.
It is clear that the Jacobian of this map is nonzero. Then inverting
this map, we can regard $x(s,t), y(s,t)$ as functions of $s,t$
(or as power series in $q_1:=e^s, q_2=e^t$). We wish
to derive a system of polynomial PDEs for these functions. Recall the
transformation laws under a change of variables
\eqn\dumb{\eqalign{
\partial_x f=&~{[y,f]\over [y,x]}\cr
\partial_y f=&~{[x,f]\over [x,y]} }}
where $[f,g]$ is the ``Poisson bracket'':
\eqn\dumb{[f,g]:=\partial_s f\partial_t g-\partial_s g\partial_t f.}
Under this change of variables, $L_1, L_2$ becomes
\eqn\changevar{\eqalign{
\cL_1=&\sum_{0\leq i+j\leq m} c_{ij}\partial_s^i\partial_t^j\cr
\cL_2=&\sum_{0\leq i+j\leq2} d_{ij}\partial_s^i\partial_t^j. }}
{}From the transformation laws, it is easy to see that
up to an overall factor the new coefficients
$c_{ij},d_{ij}$ are differential {\it polynomials} of $x(s,t),y(s,t)$.

Now observe that
\eqn\shift{\eqalign{
\cL_1 (s w_0)- s \cL_1 w_0 =&~
\sum i c_{ij}\partial_s^{i-1}\partial_t^jw_0=0\cr
\cL_1 (t w_0)- t \cL_1 w_0 =&~
\sum j c_{ij}\partial_s^i\partial_t^{j-1}w_0=0\cr
\cL_2 (s w_0)- s \cL_2 w_0 =&~
d_{10}w_0+d_{11} \partial_t w_0 +2d_{20}\partial_s w_0 =0\cr
\cL_2 (t w_0)- t \cL_2 w_0 =&~
d_{01}w_0+d_{11} \partial_sw_0 +2d_{02}\partial_t w_0 =0\cr
\cL_1w_0=&~0\cr
\cL_2w_0=&~0.}}
It is easy to check that $(d_{11}^2-4d_{02}d_{20})[x,y]^2=
b_{11}^2-4b_{02}b_{20}$ which is nonzero. Thus we can solve for
$\partial_s w_0, \partial_t w_0$ in terms of $w_0$ and differential
expressions in $x,y$ in the third and fourth equations in \shift.
We can hence use this to eliminate all
higher derivatives $\partial_s^i\partial_t^jw_0$ in the first and second
equations
in \shift. But since $w_0$ appears linearly everywhere,
we can factor it out and obtain a pair of coupled polynomial PDEs
in $x,y$. Their order is at most $m$. Thus we have
\proposition{Given the linear PDE system $L_1,L_2$ above, there exist
a pair of polynomial PDEs for $x,y$.}

Note that the system \shift~ can be regarded as a (overdetermined) system
of polynomial PDEs in $x,y,w_0$. We will use this system to
study the mirror map and the Yukawa couplings by means of
perturbation theory in the next section.

As an example, consider the system \generalPDE~ which we can write in
the form \generaltwo~, where $m=2$, hence \changevar~ under a change of
variables.
As we have seen, such a system arises as the Picard-Fuchs system
for certain 2-moduli families of toric K3 surfaces.
The hypotheses on the uniqueness of series solutions,
and that $b_{11}^2-4b_{02}b_{20}$ is nonzero  can be
easily checked.
In this case our polynomial PDEs for $x(s,t),y(s,t)$ becomes
\eqn\dumb{\eqalign{
(b_{11}^2-4b_{02}b_{20})(2c_{02}c_{10}-c_{01}c_{11})=&~
(a_{11}^2-4a_{02}a_{20})(2d_{02}d_{10}-d_{01}d_{11})\cr
(b_{11}^2-4b_{02}b_{20})(2c_{20}c_{01}-c_{10}c_{11})=&~
(a_{11}^2-4a_{02}a_{20})(2d_{20}d_{01}-d_{10}d_{11}). }}

\newsec{Perturbations}
\seclab\perturbations

We have seen a
 nice description of the periods and the mirror map when one degenerates
a certain family of Calabi-Yau threefolds
along a codimension one subfamily.
We would like to use perturbation theory to study the Calabi-Yau threefolds
in a neighborhood of the codimension one subfamily
Consider for example the mirror threefold $X$
 of the degree 12 hypersurfaces in
$\P^4[6,2,2,1,1]$ which has $h^{1,1}=2$. Let $x,y$ be the Mori coordinates
near the large volume limit of $X$. It is now known that the periods and
the mirror map
in this case admit a description in terms of the $j$ function
in the limit $y\rightarrow 0$. Can one give a similar description
near $y=0$? How about near $x=0$?

We will give two descriptions in the case of the
degree 12 hypersurface above.
We show that mirror map and the Yukawa couplings, order by order in $q_2$,
 can be described by quadrature
in terms of the $j$ function. In fact, we will
compute the first few terms. The second description is  by means of
perturbation theory in the $q_1$ direction. Here remarkably, the first
few terms are purely algebraic, rather than transcendental.

We briefly review what is known for the degree 12 hypersurfaces in
$\P^4[1,1,2,2,6]$. The Picard-Fuchs system in
the Mori coordinates $x,y$ is given by
\eqn\pfsystem{\eqalign{
L_1=&~\Theta_x^2(\Theta_x-
2\Theta_y)-8x(6\Theta_x+5)(6\Theta_x+3)(6\Theta_x+1)\cr
L_2=&~\Theta_y^2-y(2\Theta_y-\Theta_x+1)(2\Theta_y-\Theta_x).}}
This motivates the study of the following family of PDEs, where $\lambda,\nu$
are constants,
(cf. \lianyaufour~ and see also Appendix B):
\eqn\pfgeneral{\eqalign{
L_1=&~\Theta_x^2(\Theta_x-
2\Theta_y)-\lambda x(\Theta_x+1/2)(\Theta_x+1/2-\nu)(\Theta_x+1/2+\nu)\cr
L_2=&~\Theta_y^2-y(2\Theta_y-\Theta_x+1)(2\Theta_y-\Theta_x).}}
As before there are unique solutions near $x=y=0$ of the form
$w_0=1+O(h), w_1=w_0Log~x+O(h), w_2=w_0Log~y+O(h)$, and the
coefficients of $L_2$ satisfies $b_{11}^2-4b_{02}b_{20}=-4y^2\neq0$.
Thus the system \pfsystem~ is of the type \generaltwo~ with $m=3$.
Associated to it is the nonlinear system \shift.
The ``mirror map''
$(t_1,t_2)\mapsto(x(q_1,q_2), y(q_1,q_2))$
is defined locally by the inverse of the power series relations
\eqn\dumb{\eqalign{
q_1=&~e^{w_1(x,y)/w_0(x,y)}=x(1+O(h))\cr
q_2=&~e^{w_2(x,y)/w_0(x,y)}=y(1+O(h)).}}
Thus we can write
\eqn\xyw{\eqalign{
x(q_1,q_2)=&~\sum_{i=0}^\infty x_i(q_1) q_2^i\cr
y(q_1,q_2)=&~\sum_{i=1}^\infty y_i(q_1) q_2^i\cr
w_0(x(q_1,q_2),y(q_1,q_2))=&~\sum_{i=0}^\infty g_i(q_1) q_2^i\cr
}}
where the $x_i(q_1), y_i(q_1), g_i(q_1)$ are power series. We will use the PDEs
\shift~ derived in the last section to compute
the $x_i, y_i, g_i$, $i\leq2$. The results turns out to have a universal form.

\theorem{For $x_i, y_i, g_i$ as defined in \xyw, $x_0$ is the unique power
series solution with $x_0=q_1+O(q_1^2)$ to the Schwarzian equation
$2Q(x_0)x_0'^2+\{x_0,s\}=0$ with
$Q(x)={1+(-{5\over4}+\nu^2)\lambda x +(1-\nu)(1+\nu)\lambda^2 x^2\over
4 x^2 (1-\lambda x)^2}$. We also have
\eqn\desired{\eqalign{
g_0=&~{x_0'\over x_0(1-\lambda x_0)^{1/2}}\cr
x_1=&~{-x_0^2  y_1\over g_0x_0'^2}(-2 g_0' x_0' + g_0 x_0'')\cr
y_1=&~exp\left(2\int\int\int
{(Log~g_0)' (Log~x_0)''-(Log~g_0)'' (Log~x_0)'-(Log~g_0)'^2 (Log~x_0)'\over
1-\lambda x_0} ds ds ds\right)\cr
g_1=&~{1\over x_0'^3}(
x_1 g_0' x_0'^2  + x_0^2  y_1 x_0' g_0'' -
x_0^2  y_1 g_0' x_0'')}}
where prime here means $\Theta_{q_1}$.
}\thmlab\fourone

%%\comment{See computer file "eqn", univ1,univ2,univ3.}

\proof
For the proof of the statements concerning $x_0,g_0$, see \lianyaufour.
We will study the PDE system \shift~ associated to \pfsystem~ up to first order
in powers of $q_2=e^t$.
We substitute $x=x_0+q_2 x_1 +O(q_2^2)$,
$y=q_2 y_1+ O(q_2^2)$, $w_0=g_0+q_2 g_1+O(q_2^2)$ into \shift~ where the
$x_i,y_i,g_i$
are all power series in $q_1=e^s$.
To lowest order,
the first equation in \shift~ is a polynomial ODE
in $g_0,x_0$. Using the result \lianyaufour~ that
$w_0(x(q_1),0)^2=g_0^2={x_0'^2\over x_0^2(1-1728x_0)}$,
it is easy to show that this ODE holds identically.
Now consider the second equation in \shift.
To leading order it is a complicated polynomial ODE in $y_1,g_0, x_0$.
But after we apply repeatedly the fact that $x_0$ satisfies the Schwarzian
equation to eliminate $x_0''', x_0'''',...$, we see that the equation
is solvable. The general solution is exactly
%\eqn\dumb{\eqalign{
%& -y_1^3  x_0'^4  + 6672 x_0 y_1^3  x_0'^4  -
%  10782720 x_0^2  y_1^3  x_0'^4  - 2 x_0^3  x_0' y_1'^3  +\cr&~
%  10368 x_0^4  x_0' y_1'^3 - 17915904 x_0^5  x_0' y_1'^3  +
%  10319560704 x_0^6  x_0' y_1'^3  +
%2 x_0 y_1^2  x_0'^2  x_0'' - \cr &~
% 10368 x_0^2  y_1^3  x_0'^2  x_0'' +
% 11943936 x_0^3  y_1^3  x_0'^2  x_0'' - x_0^2  y_1^3  x_0''^2  +
%  3456 x_0^3 y_1^3  x_0''^2  - \cr &~
%2985984 x_0^4  y_1^3  x_0''^2  +
%   3 x_0^3  y_1 x_0' y_1' y_1'' -
%  15552 x_0^4  y_1 x_0' y_1' y_1'' +
% 26873856 x_0^5  y_1 x_0' y_1' y_1'' - \cr &~
%  15479341056 x_0^6  y_1 x_0' y_1' y_1'' -
% x_0^3  y_1^2  x_0' y_1'''    + 5184 x_0^4  y_1^2  x_0' y_1'''    -
%   8957952 x_0^5  y_1^2  x_0' y_1'''    + \cr &~
%   5159780352 x_0^6  y_1^2  x_0' y_1''' =0.}}
the third equation in \desired. There is a unique particular solution
which is a power series in $q_1$ with leading coefficient 1.
Similarly the third and fifth equations in \shift, to lowest order, gives
the second and fourth equations in \desired. \qed

We can also use the same system \shift~ to compute the higher order terms.
At each order $i\geq 2$ the triple $(x_i,y_i,g_i)$, can now
be solved successively
in terms of the lower terms by applying the third, fourth and sixth
equations in \shift. It appears that at each order $i\geq2$, $x_i,y_i,g_i$
are given by some differential {\it rational} functions of the lower order
terms without solving a differential equation. For example, $x_2,y_2,g_2$
in fact occurs linearly (with no derivative thereof)
in the following equations. They are in fact the order $O(q_2^2)$ terms
of the third, fourth and sixth equations respectively in \shift:
\eqn\dumb{\eqalign{
 &~-2 x_1 y_1 g_0' x_0'^2  + 4 x_0 y_1^2  g_0' x_0'^2  +
 2 g_1 y_1 x_0'^3  - 2 g_0 y_1^2  x_0'^3  - \cr&~
 g_0 y_2 x_0'^3  +
 2 x_0^2  y_1 g_0' x_0' y_1' + g_0 x_1 x_0'^2  y_1' -
 g_0 x_0^2  x_0' y_1'^2  - \cr&~
 g_0 x_0^2  y_1 y_1' x_0'' +
 g_0 x_0^2  y_1 x_0' y_1''=0\cr
&\cr
&~ 2 x_1^2  y_1 g_0' x_0'^2  -
 12 x_0 x_1 y_1^2  g_0' x_0'^2  -
  2 x_0^2  y_1 y_2 g_0' x_0'^2  -
 2 x_0^2  y_1^2  g_1' x_0'^2  -  \cr&~
  3 g_1 x_1 y_1 x_0'^3  - 4 g_0 x_2 y_1 x_0'^3  +
  4 g_1 x_0 y_1^2  x_0'^3  + 6 g_0 x_1 y_1^2  x_0'^3  +  \cr&~
  3 g_0 x_1 y_2 x_0'^3  + 4 x_0^2  y_1^2  g_0' x_0' x_1' +
   3 g_0 x_1 y_1 x_0'^2  x_1' -
   4 g_0 x_0 y_1^2  x_0'^2  x_1' -  \cr&~
    4 x_0^2  x_1 y_1 g_0' x_0' y_1' -
   3 g_0 x_1^2  x_0'^2  y_1' + 2 g_1 x_0^2  y_1 x_0'^2  y_1' +  4 g_0 x_0 x_1
y_1 x_0'^2  y_1' -  \cr&~
   2 g_0 x_0^2  y_1 x_0' x_1' y_1' +
    2 g_0 x_0^2  x_1 x_0' y_1'^2  -
    g_0 x_1^2  y_1 x_0' x_0'' +
g_1 x_0^2  y_1^2  x_0' x_0'' +  \cr&~
 6 g_0 x_0 x_1 y_1^2  x_0' x_0'' +
 g_0 x_0^2  y_1 y_2 x_0' x_0'' -
   3 g_0 x_0^2  y_1^2  x_1' x_0'' +
    3 g_0 x_0^2  x_1 y_1 y_1' x_0'' +  \cr&~
    g_0 x_0^2  y_1^2  x_0' x_1'' -
    g_0 x_0^2  x_1 y_1 x_0' y_1''=0\cr
 &\cr
&~-4 x_2 y_1 g_0' x_0'^3  + 6 x_1 y_1^2  g_0' x_0'^3  +
   3 x_1 y_2 g_0' x_0'^3  - 3 x_1 y_1 g_1' x_0'^3  + \cr&~
   4 x_0 y_1^2  g_1' x_0'^3  + 4 g_2 y_1 x_0'^4  -
   6 g_1 y_1^2  x_0'^4  - 3 g_1 y_2 x_0'^4  + \cr&~
    3 x_1 y_1 g_0' x_0'^2  x_1' -
    4 x_0 y_1^2  g_0' x_0'^2  x_1' -
 3 x_1^2  g_0' x_0'^2  y_1' +
   4 x_0 x_1 y_1 g_0' x_0'^2  y_1' + \cr&~
   2 x_0^2  y_1 g_1' x_0'^2  y_1' + 3 g_1 x_1 x_0'^3  y_1' -
    4 g_1 x_0 y_1 x_0'^3  y_1' -
    2 x_0^2  y_1 g_0' x_0' x_1' y_1' + \cr&~
    2 x_0^2  x_1 g_0' x_0' y_1'^2  -
2 g_1 x_0^2  x_0'^2  y_1'^2  +
    x_1^2  y_1 x_0'^2  g_0'' -
 6 x_0 x_1 y_1^2  x_0'^2  g_0'' - \cr&~
    x_0^2  y_1 y_2 x_0'^2  g_0'' +
    2 x_0^2  y_1^2  x_0' x_1' g_0'' -
    2 x_0^2  x_1 y_1 x_0' y_1' g_0'' -
    x_0^2  y_1^2  x_0'^2  g_1'' - \cr&~
 x_1^2  y_1 g_0' x_0' x_0'' +
  6 x_0 x_1 y_1^2  g_0' x_0' x_0'' +
    x_0^2  y_1 y_2 g_0' x_0' x_0'' +
    x_0^2  y_1^2  g_1' x_0' x_0'' - \cr&~
    3 x_0^2  y_1^2  g_0' x_1' x_0'' +
    3 x_0^2  x_1 y_1 g_0' y_1' x_0'' -
    g_1 x_0^2  y_1 x_0' y_1' x_0'' +
   x_0^2  y_1^2  g_0' x_0' x_1'' - \cr&~
    x_0^2  x_1 y_1 g_0' x_0' y_1'' +
    g_1 x_0^2  y_1 x_0'^2  y_1''=0}}
%%\comment{Computer files "eqn", univ4,univ5,univ6.}
We have also checked that the order $O(q_2^3)$ term of the
third, fourth and sixth equations in \shift are linear in
$x_3,y_3,g_3$ (with no derivative thereof), which determines
this triple in terms of the lower order terms. We emphasize that
the ODEs above are universal in sense that they are independent
of the parameters $\lambda,\nu$ of our linear PDEs.
%%\comment{Computer files "eqn", eqn8,eqn9,eqn10.}

\corollary{For the case of the degree 12 hypersurface in $\P^4[6,2,2,1,1]$,
the $x_i, y_i, g_i$, $i\leq3$, are given in terms of the $j$ function
by quadrature.}

\proof In this case, we have $(\lambda,\nu)=(1728,1/3)$ and we check
that the unique solution to the Schwarzian equation $2Qx_0'^2+\{x_0,s\}=0$
is given by $x_0=1/j$. By theorem above and the remarks following it,
we see that the $x_i, y_i, g_i$, $i\leq3$ can be expressed (explicitly!)
in terms of $j$ by quadrature. \qed

\subsec{remarks}
\seclab\remarks

1. When $(\lambda,\nu)=(1728,1/3)$, if we use the Schwarzian equation for $j$
to further simplifies things, we get
\eqn\desired{\eqalign{
x_0=&~1/j\cr
g_0=&~E_4^{1/2}\cr
x_1=&~y_1{x_0(2x_0'^2-5184 x_0x_0'-x_0x_0''+1728x_0^2x_0'')\over
(1-1728x_0)x_0'^2}\cr
y_1=&~exp\left(\int\int\int
-((1-1728 x_0)^2 x_0^2 x_0''^2 -
2(1-3456 x_0)(1-1728 x_0) x_0 x_0'^2 x_0'' +\right.\cr
&~\left.(1-6672 x_0+10782720 x_0^2) x_0'^4)/(x_0^3 (1-1728 x_0)^3 x_0')
ds ds ds\right)
}}

2. It turns out that in the example above, the $O(q_2^k)$ coefficient $g_k$ of
$w_0(x(q_1,q_2), y(q_1,q_2))$ is always a rational function of the
lower order coefficients for $x,y$. More precisely, we have:

\lemma{For each $k$, $g_k$ as a power series is a rational function of
$g_0, x_0,x_0',x_0'',$ $x_1,..,x_k, y_1,..,y_k$, which is polynomial in
$x_1,..,x_k, y_1,..,y_k$.}
\thmlab\fourtwo

\proof Since $g_0^2={x_0'^2\over x_0^2(1-1728x_0)}$ and since $x_0=1/j$
satisfies a third order Schwarzian equation, a
rational function in $g_0,g_0',...,x_0,x_0',...$
can be reduced to one in $g_0, x_0,x_0',x_0''$. It is enough then to
show that ${\partial^k w_0\over\partial q_2^k}|_{q_2=0}= k! g_k$
lives in the ring:
\eqn\dumb{
\cR=\C(g_0,g_0',...,x_0,x_0',...)[x_1,..,x_k,y_1,..,y_k].}

Observe that
\eqn\wc{\eqalign{
w_0(x,y)=&~\sum_{n\geq2m\geq0}c(n,m)x^ny^m
=\sum_{m\geq0} {1\over m!^2}y^m x^{2m}f^{(2m)}(x)~~{\sl where}\cr
c(n,m):=&~{(6n)!\over (3n)! n!^2 m!^2 (n-2m)!}\cr
f(x):=&~\sum_{n\geq0}c(n,0)x^n\cr
g_0=&~f(x_0(q_1)).}}
It follows that ${\partial^k w_0\over\partial q_2^k}|_{q_2=0}$ is a sum of
terms of the form
${\partial^a y^m \over\partial q_2^a}
{\partial^b x^{2m}\over\partial q_2^b} $
${\partial^c f^{(2m)}(x) \over\partial q_2^c}|_{q_2=0}$,
with $0\leq a,b,c\leq k$. But
${\partial^a y^m \over\partial q_2^a}|_{q_2=0}$ is zero for all $m>a$, because
$y^m=q_2^m(1+O(h))$, and is a polynomial of $y_1,..,y_k$ for $m\leq a\leq k$,
hence is in $\cR$.
Similarly, the ${\partial^b x^{2m} \over\partial q_2^b} |_{q_2=0}$ are
polynomials of $x_0,x_1,..,x_k$, hence are in $\cR$.
 Finally (by ${\partial \over\partial q_2}=
{\partial x\over\partial q_2}{d\over dx}$, $x|_{q_2=0}=x_0$) the
${\partial^c f^{(2m)}(x) \over\partial q_2^c}|_{q_2=0}$
are clearly polynomials in
$x_0,x_1,..,x_k$ and $f(x_0),f'(x_0),...$.
But since $g_0=f(x_0(q_1))$ and $f'(x_0)=g_0'/x_0'\in\C(g_0,x_0,x_0')$, it
follows that these polynomials are also in $\cR$. \qed

3. One of the consequences of the fact that the
restriction of the mirror map is given
by the $j$ function is that the mirror map cannot be algebraic. More
precisely, there is no nontrivial polynomial relations
\eqn\dumb{\eqalign{
P(x,y,q_1,q_2)=&~0\cr
Q(x,y,q_1,q_2)=&~0}}
along the graph of the mirror map $(q_1,q_2)\mapsto(x,y)$. To see this,
suppose both $P,Q$ are irreducible. Then from the resultants of the two
polynomials we obtain two irreducible polynomial relations, along the graph:
\eqn\dumb{\eqalign{
\tilde{P}(x,q_1,q_2)=&~0\cr
\tilde{Q}(y,q_1,q_2)=&~0.}}
By irreduciblity, the polynomial in two variables $\tilde{P}(a,b,0)$ is not
{\it identically} zero. But
under the mirror map we have $(q_1,0)\mapsto(1/j(q_1),0)$, implying that
$\tilde{P}(1/j(q_1),q_1,0)=0$ identically, which is absurd.

However it turns out that the mirror map in this case is very close to being
algebraic in the sense we shall explain in the next section.

4. What we have effectively described above is a perturbation method for
computing, order by order in one of the Kahler coordinates $q_2$,
the mirror map given by $x(q_1,q_2), y(q_1,q_2)$, and the
special period given by $w_0$. The perturbation series
\xyw~ turn out to be useful also for computing the Yukawa couplings via
mirror symmetry. Recall that the type A couplings of
a Calabi-Yau variety $X$ is given in terms of the type B coupling
of the mirror variety $Y$ via the formulas \cdgp\HKTYI\COFKM:
\eqn\KA{
K^A_{ijk}={1\over w_0^2}\sum_{l,m,n}
{\partial x_l\over\partial t_i}
{\partial x_m\over\partial t_j}
{\partial x_n\over\partial t_k} K^B_{lmn}(x) }
where the $K^B$ are rational functions.
In the 2-moduli example
above, these rational functions have been computed in \HKTYI (up to
multiplicative constants):
\eqn\dumb{\eqalign{
K^B_{1,1,1}=&~{4\over \bar{x}^3 D}\cr
K^B_{1,1,2}=&~{2(1-\bar{x})\over \bar{x}^2\bar{y} D}\cr
K^B_{1,1,1}=&~{2\bar{x}-1\over \bar{x}\bar{y}(1-\bar{y})D}\cr
K^B_{1,1,1}=&~{1-\bar{x}+\bar{y}-3\bar{x}\bar{y}\over
2\bar{y}^2(1-\bar{y})^2 D} }}
where $D=(1-\bar{x})^2-\bar{x}^2\bar{y}$, $\bar{x}=1728 x,\bar{y}=4y$.
It follows that in this case
the $K^A$ are computable order by order in terms of modular functions
simply by computing the $x,y,w_0$ using the perturbation method above.
Up to order $O(q_2^2)$, the answers are given in Appendix A.

Now on the other hand
the type A coupling takes the form
\eqn\dumb{
K^A_{ijk}=K^0_{ijk}+\sum_{d_1+d_2>0}
{n_{d_1,d_2} d_i d_j d_k q_1^{d_1}q_2^{d_2}\over 1-q_1^{d_1}q_2^{d_2} } }
where the $K^0_{ijk}$ are the classical cubic intersection numbers  of
$X$, and the $n_{d_1,d_2}$ is the mirror symmetry prediction for the number
of rational curves of degrees $(d_1,d_2)$. If we write
\eqn\dumb{
K^A_{ijk}=\sum_{m=0}^\infty K_{ijk}[m]q_2^m}
where the $K_{ijk}[m]$ are power series in $q_1$, then using ($m>0$)
\eqn\dumb{
{\partial^m\over\partial q_2^m}|_{q_2=0}
{q_1^{d_1}q^{d_2}\over 1-q_1^{d_1}q_2^{d_2} }
=\left<\matrix{ m! q_1^{d_1m\over d_2} & {\sl if~}d_2|m\cr
0 &{\sl otherwise} }\right.
}
it is easy to show that
\eqn\dumb{\eqalign{
K_{ijk}[0]=&~K^0_{ijk}+\delta_{i,1}\delta_{j,1}\delta_{k,1}\sum_{d_1>0}
{n_{d_1,0} d_1^3q_1^{d_1}\over 1-q_1^{d_1} }\cr
K_{ijk}[m]=&~\sum_{d_1\geq0, d_2|m}
n_{d_1,d_2} d_i d_j d_k q_1^{d_1m\over d_2} ~~{\sl for~}m>0. }}
Thus using perturbation theory each of the $K_{ijk}[m]$ can now
be expressed in terms of modular functions.

\newsec{Perturbations Around $x=0$}
\seclab\perturbationx

We now interchange the roles of $(x,q_1)$ and $(y,q_2)$. One
might expect that the discussion above would carry over with few
changes. It turns out that while all the techniques carry over,
the results have vast simplifications in this case.
This consideration is motivated by a few observations.

First note that the Picard-Fuchs system \pfsystem~ is highly asymmetric
in $x,y$. Thus its reasonable that the two limits along $y=0$ and
$x=0$ are qualitatively different. Second note that along $x=0$,
the solutions $w_0,w_2$ degenerate to elementary functions
\eqn\dumb{\eqalign{
w_0(0,y)=&~1\cr
w_2(0,y)=&~Log(1-\sqrt{1-4y}-2y)-Log(2y), }}
and they are solutions to $\Theta_y^2-2y(2\Theta_y+1)\Theta_y$.
It is then easy to compute the mirror map restricted along $x=0$:
$y(0,q_2)={q_2\over(1+q_2)^2}$ which is rational rather than
transcendental!
Third from the definition of the series $x,y,w_0$, we can write
\eqn\xywtwo{\eqalign{
x(q_1,q_2)=&~\sum_{i=1}^\infty X_i(q_2) q_1^i\cr
y(q_1,q_2)=&~\sum_{i=0}^\infty Y_i(q_2) q_1^i\cr
w_0(x(q_1,q_2),y(q_1,q_2))=&~\sum c(n,m) x(q_1,q_2)^ny(q_1,q_2)^m
=\sum_{i=0}^\infty G_i(q_2) q_1^i
}}
where the $X_i, Y_i, G_i$ are power series.
Let's describe the $G_i$ in terms of the $X_i,Y_i$.
The $c(n,m)$ are
such that $c(n,m)=0$ for all $2m>n$. (The same argument below
applies to any other 2-moduli family of
Calabi-Yau toric varieties with fundamental period having the form
$w_0(x,y)=\sum c(n,m)x^ny^m$ such that for each $n$, $c(n,m)=0$ for $m>>n$;
see the Appendix B for more examples.)
Since $x(q_1,q_2)^n=q_1^n(1+O(h))$ and since $c(n,m)=0$ for $m>>n$,
at most finitely many terms in the sum $\sum c(n,m) x(q_1,q_2)^ny(q_1,q_2)^m$
contribute to a given $G_k$. Thus it is a {\it finite}
linear sum of $X_iY_j$. In particular, if the $X_i,Y_j$ are
algebraic then so are the $G_k$. As seen above,
\eqn\dumb{\eqalign{
G_0=&~1\cr
Y_1=&~{q_2\over(1+q_2)^2}. }}

In fact applying perturbation theory along the $q_1$ direction
 on the PDE system \shift, we compute the first few terms.
The emerging pattern is clear evidence that the $X_i,Y_i,G_i$ are in fact
rational.

\proposition{Denote $q:=q_2$. Then
\eqn\dumb{\eqalign{
G_0=&~1\cr
G_1=&~120(1+q)\cr
G_2=&~360 (-17 + 268 q - 17 q^2 )\cr
G_3=&~480 (1 + q) (1537 + 135866 q + 1537 q^2 )\cr
G_4=&~120 (-893747 + 362384432 q + 1610384580 q^2  +
 362384432 q^3  - 893747 q^4 )\cr
G_5=&~720 (1 + q) (24145921 + 38170176314 q +
 411770251626 q^2  + 38170176314 q^3  + 24145921 q^4 )\cr
&\cr
X_1=&~1+q\cr
X_2=&~-24 (31 + 82 q + 31 q^2 )\cr
X_3=&~36 (1 + q) (9907 + 6130 q + 9907 q^2 )\cr
X_4=&~-64 (2193143 + 8342176 q + 9151506 q^2  + 8342176 q^3  +
      2193143 q^4 )\cr
X_5=&~30 (1 + q) (1644556073 - 1014171566 q -
26082465678 q^2  - 1014171566 q^3  + 1644556073 q^4 )\cr
&\cr
Y_0=&~q/(1+q)^2\cr
Y_1=&~-240 q (1-q)^2/(1+q)^3\cr
Y_2=&~-360 q(1 - q)^2  (37 + 554 q + 37 q^2 )/(1+q)^4\cr
Y_3=&~-320 q(1 - q)^2
(7747 + 393600 q + 1117306 q^2  +
393600 q^3  + 7747 q^4)/(1+q)^5\cr
Y_4=&~-60 q(1 - q)^2
(14352887 + 1931431324 q + 10227963073 q^2  + 17727689272 q^3  +\cr&~
10227963073 q^4  + 1931431324 q^5  + 14352887 q^6 )/(1+q)^6. }}
}
\thmlab\fiveone

\proof $X_i,Y_i,G_i$, $i\leq3$, can be computed by solving the differential
equations \shift~ order by order as we have done before.
But this will be hard without first knowing the answers. So we use
the following slightly different approach. Numerically it is easy
to compute $x,y,w_0$ as a power series in $q_1,q_2$ up total order
say $O(h^{15})$.
We first guess an ansatz (the list above) for the
$X_i,Y_i,G_i$ based on the numerical results.
Then we check that our ansatz satisfies our differential equations
derived from \shift~ (up to $O(q_2^6)$) governing the $X_i,Y_i,G_i$.
Observe also
that the ODEs for the $X_i,Y_i,G_i$
derived from \shift~
 can have three as the highest order in derivatives. This
means that the differential equations together with the first three
coefficients of each of the $X_i,Y_i,G_i$ determines the whole series
$X_i,Y_i,G_i$ uniquely. The first three coefficients of our ansatz
are easily check to be correct. \qed

Clearly we can apply our perturbation argument to the Yukawa couplings
$K^A_{ijk}$ as in the previous section, with the roles of $q_1,q_2$
interchanged. Thus we can write
\eqn\dumb{
K^A_{ijk}=\sum_{m=0}^\infty K_{ijk}[m]q_1^m}
where now the $K_{ijk}[m]$ are power series in $q_2$.
Then
\eqn\dumb{\eqalign{
K_{ijk}[0]=&~K^0_{ijk}+\delta_{i,2}\delta_{j,2}\delta_{k,2}\sum_{d_2>0}
{n_{0,d_2} d_2^3q_2^{d_2}\over 1-q_2^{d_2} }\cr
K_{ijk}[m]=&~\sum_{d_2\geq0, d_1|m}
n_{d_1,d_2} d_i d_j d_k q_2^{d_2m\over d_1} ~~{\sl for~}m>0. }}
If the $X_i,Y_i,G_i$, for $i\leq l$, are rational, then so are the
couplings $K_{ijk}[m]$, for $m\leq l$. The proposition above shows that
this is true for at least small $l$. In fact using the proposition above,
we have

\eqn\dumb{\eqalign{
K^A_{111}=&~ 4 + 2496 (1 + q_2) q_1 +
1152 (1556 + 13481 q_2 + 1556 q_2^2 ) q_1^2  + \cr&~
4768 (1358353 + 46666143 q_2 + 46666143 q_2^2  + 1358353 q_2^3 ) q_1^3  +
O(q_1^4)\cr
K^A_{112}=&~ 2 + 2496 q_2 q_1 + 576 q_2 (13481 + 3112 q_2) q_1^2 + \cr&~
 768 q_2 (15555381 + 31110762 q_2 + 1358353 q_2^2 ) q_1^3  + O(q_1^4)\cr
K^A_{122}=&~  2496 q_2 q_1 + 288 q_2 (13481 + 6224 q_2) q_1^2  + \cr&~
 768 q_2 (5185127 + 20740508 q_2 + 1358353 q_2^2 ) q_1^3 + O(q_1^4)\cr
K^A_{222}=&~{2 q_2\over 1-q_2} +
2496 q_2 q_1 + 144 q_2 (13481 + 12448 q_2) q_1^2  + \cr&~
256 q_2 (5185127 + 41481016 q_2 + 4075059 q_2^2 ) q_1^3  + O(q_1^4). }}

These formulas give  infinitely many $n_{d_1,d_2}$ simultaneously!
For example, we have
\eqn\dumb{
n_{0,d_2}=2\delta_{d_2,1}.}
For another example, for at least $d_1=0,1,..,4$,
we have $n_{d_1,d_2}=0$ for all but finitely many $d_2$. The nonzero ones
can be computed immediately from the formulas above.

{\bf Acknowledgements:} We thank A. Kachru, A. Klemm, S. Theisen and C. Vafa
for helpful discussions.

\newsec{Appendix A}

In this appendix, we compute perturbatively the type A Yukawa couplings
$K^A_{ijk}$ (see discussions in section \remarks) for the family of
Calabi-Yau toric varieties corresponding to the degree
12 hypersurface in $\P^4[6,2,2,1,1]$.
We perturb in the neighborhood
of the codimension one subfamily with $y=0$ up to $O(y^2)$ or equivalently
$O(q_2^2)$. We give $K^A_{ijk}[0], K^A_{ijk}[0]$, as differential
rational functions in $x_i,y_i,g_i$, $i\leq2$, which we have given
explicitly in terms of the $j$ function by quadrature in section
\perturbations.
The computation for higher order terms is straightforward but tedious.
$K^A_{111}[0], K^A_{112}[0]$ have already been considered in \vk\lkt.
\vfill\eject

\eqn\dumb{\eqalign{
K^A_{111}[0]=&~{4x_0'\over x_0(1-1728x_0)}+6{y_1'\over y_1}\cr
K^A_{111}[1]=&~ ({-2 g_1\over g_0^3  (1 - 1728 x_0)^2}+
 {3456 (x_1 - 1728 x_0 x_1 + 3456 x_0^2  y_1)\over
  g_0^2  (-1 + 1728 x_0)^4})({4 x_0'^3\over x_0^3}+
{6 (1 - 1728 x_0) x_0'^2  y_1'\over x_0^2  y_1})+\cr&~
({12 x_0'^2  (-(x_1 x_0') + x_0 x_1')\over  x_0^4}+
{ 12 (-1 + 3456 x_0) x_0' y_1'^2\over x_0 y_1}+
{ 2 (1 - 1728 x_0) y_1'^3\over y_1^2}+\cr&~
 {12 (-1 + 864 x_0) x_0'^2 x_1y_1'\over x_0^3  y_1}+
 { 6(-1 + 1728 x_0) x_0'^2y_2y_1'\over x_0^2  y_1^2}+
{12 (1 - 1728 x_0) x_0' x_1'y_1'\over x_0^2  y_1}+\cr&~
{6 (1 - 1728 x_0) x_0'  y_2'\over x_0^2  y_1})/
 (g_0^2  (1 - 1728 x_0)^2 ) \cr
K^A_{112}[0]=&~2\cr
K^A_{112}[1]=&~
 2 (2 g_1 y_1 x_0'^2  -
 6912 g_1 x_0 y_1 x_0'^2  +
   5971968 g_1 x_0^2  y_1 x_0'^2  -
   5184 g_0 x_1 y_1 x_0'^2  + \cr&~
  8957952 g_0 x_0 x_1 y_1 x_0'^2  -
   11943936 g_0 x_0^2  y_1^2  x_0'^2  -
  g_0 y_2 x_0'^2  +
   3456 g_0 x_0 y_2 x_0'^2  - \cr&~
  2985984 g_0 x_0^2  y_2 x_0'^2  -
  2 g_0 y_1 x_0' x_1' +
    5971968 g_0 x_0^2  y_1 x_0' x_1' -
  2 g_0 x_1 x_0' y_1' + \cr&~
     6912 g_0 x_0 x_1 x_0' y_1' -
    5971968 g_0 x_0^2  x_1 x_0' y_1' +
    4 g_0 x_0 y_1 x_0' y_1' -
    20736 g_0 x_0^2  y_1 x_0' y_1' + \cr&~
  23887872 g_0 x_0^3  y_1 x_0' y_1' -
    g_0 x_0^2  y_1'^2  + 3456 g_0 x_0^3  y_1'^2  -
     2985984 g_0 x_0^4  y_1'^2 )/
  (g_0^3  x_0^2  (-1 + 1728 x_0)^3  y_1) \cr
K^A_{122}[0]=&~0\cr
K^A_{122}[1]=&~ 2 (-2 x_1 x_0' + 3456 x_0 x_1 x_0' +
  2 x_0 y_1 x_0' - 6912 x_0^2  y_1 x_0' - \cr&~
 x_0^2  y_1' + 1728 x_0^3  y_1')) /
 ((-1 + 1728 x_0) x_0'^2 )\cr
K^A_{222}[0]=&~0\cr
K^A_{222}[1]=&~{2 y_1 \over   g_0^2  (1 - 1728 x_0)}
}}

\newsec{Appendix B}

In this appendix, we will study using perturbation technique introduced
above, the perturbations around $y=0$ ($q_2=0$) and $x=0$ ($q_1=0$) for
the following five families Calabi-Yau
toric varieties, all of which have $h^{1,1}=2$:
\eqn\dumb{\eqalign{
a.&~ {\sl degree~8~in~}\P^4[2,2,2,1,1]\cr
b.&~ {\sl degrees~(6,4)~in~}\P^5[2,2,2,2,1,1]\cr
c.&~ {\sl degrees~(4,4,4)~in~}\P^6[2,2,2,2,2,1,1]\cr
d.&~ {\sl degrees~12~in~}\P^4[4,3,2,2,1]\cr
e.&~ {\sl degrees~14~in~}\P^4[7,2,2,2,1].}}

  Throughout we will use the same notations as in our previous discussion
above.
We can use the Picard-Fuchs systems and their associated
polynomial PDEs to compute the first few
 coefficients $x_i,y_i,g_i$ of the perturbation series of the mirror map
$(q_1,q_2)\mapsto(x,y)$ and the fundamental period $w_0$ around $q_2=0$
(cf. Theorem \fourone). It turns out that theorem \fourone~
also covers the cases a, b, c. Hence all the formulas proved in that section
apply here. We won't go into the details of cases d, e,
which are a bit more tedious.
We will give the coefficients $X_i,Y_i,G_i$ of the perturbation series
around $q_1=0$ (cf. Proposition \fiveone). Note that one can now also
compute the type A Yukawa coupling $K^A_{ijk}$ order by order near
either  $q_2=0$ or $q_1=0$, by substituting
the $x_i,y_i,g_i$ (or $X_i,Y_i,G_i$) into \KA (see section \remarks).

The respective Picard-Fuchs systems for the five families above
are given by \HKTYI:
\eqn\dumb{\eqalign{
a.~~L_{1}=&~\Tx^2(\Tx-2\Ty)-4x(4\Tx+3)(4\Tx+2)(4\Tx+1)\cr
L_{2}=&~\Ty^2-y(2\Ty-\Tx+1)(\Ty-\Tx)\cr
b.~~L_{1}=&~\Tx^2(\Tx-2\Ty)-6x(2\Tx+1)(3\Tx+2)(3\Tx+1)\cr
L_{2}=&~\Ty^2-y(2\Ty-\Tx+1)(\Ty-\Tx)\cr
c.~~L_{1}=&~\Tx^2(\Tx-2\Ty)-8x(2\Tx+1)^3\cr
L_{2}=&~\Ty^2-y(2\Ty-\Tx+1)(\Ty-\Tx)\cr
d.~~L_{1}=&~\Tx^2(3\Tx-2\Ty)-36x(6\Tx+5)(6\Tx+1)
(\Ty-\Tx+2y(1+6\Tx-2\Ty))\cr
L_{2}=&~\Ty(\Ty-\Tx)-y(3\Tx-2\Ty-1)(3\Tx-2\Ty)\cr
e.~~L_{1}=&~\Tx^2(7\Tx-2\Ty)-7x(y(28\Tx-4\Ty+18)+\Ty-3\Tx-2)\cr
&~\times(y(28\Tx-4\Ty+10)+\Ty-3\Tx-1)(y(28\Tx-4\Ty+2)+\Ty-3\Tx)\cr
L_{2}=&~\Ty(\Ty-3\Tx)-y(7\Tx-2\Ty-1)(7\Tx-2\Ty) }}

As shown in \lkm~using the results of \ly, in cases a, b, c, the
 $x_0(q_1)$ are hauptmoduls
for the following genus zero groups: $\Gamma_0(2)+,\Gamma_0(3)+,\Gamma_0(4)+$.
Using a very similar argument as for Lemma \fourtwo, it is easy to show that
the lemma holds for these three cases as well. The analogue
in cases d, e are even easier because the Fourier coefficients $c(n,m)$ for
the fundamental period $w_0(x,y)$ here have the properties that
for fixed $n$ (or fixed $m$) all but finitely many $c(n,m)$ vanish.
It follows that the $g_k$ are finite sums of $x_iy_j$ in cases d, e
(cf. argument in
section \perturbationx). Similarly
in all cases, the $G_i$ are finite sums of $X_iY_j$.

\eqn\dumb{\eqalign{
case~a. & \cr
G_0 =&~ 1\cr
G_1 =&~ 24(1 + q)\cr
G_2 =&~ 24(1 + 116q + q^2)\cr
G_3 =&~ 96(1 + q)(1 + 3002q + q^2)\cr
G_4 =&~ 24(1 + 1226480q + 4864468q^2 + 1226480q^3 + q^4)\cr
G_5 =&~ 48(1 + q)(3 + 60632494q + 547690558q^2 +
     60632494q^3 + 3q^4)\cr
X_1 =&~ 1 + q\cr
X_2 =&~ -8(13 + 38q + 13q^2)\cr
X_3 =&~ 36(1 + q)(179 + 178q + 179q^2)\cr
X_4 =&~ -64(4871 + 25120q + 28658q^2 + 25120q^3 + 4871q^4)\cr
X_5 =&~ 2(1 + q)(6509415 - 12918578q - 176313170q^2 -
     12918578q^3 + 6509415q^4)\cr
Y_0 =&~ q/(1 + q)^2\cr
Y_1 =&~ -48(-1 + q)^2q/(1 + q)^3\cr
Y_2 =&~ -24(-1 + q)^2q(11 + 310q + 11q^2)/(1 + q)^4\cr
Y_3 =&~ -64(-1 + q)^2q
     (115 + 7680q + 28954q^2 + 7680q^3 + 115q^4)/(1 + q)^5\cr
Y_4 =&~ -12(-1 + q)^2q
     (37587 + 5539852q + 32302789q^2 + 62448408q^3 + \cr&~
       32302789q^4 + 5539852q^5 + 37587q^6)/(1 + q)^6}}
\eqn\dumb{\eqalign{
case~b. & \cr
G_0 =&~ 1\cr
G_1 =&~ 12(1 + q)\cr
G_2 =&~ 36(1 + 16q + q^2)\cr
G_3 =&~ 12(1 + q)(1 + 2132q + q^2)\cr
G_4 =&~ 12(7 + 94760q + 347256q^2 + 94760q^3 + 7q^4)\cr
G_5 =&~ 36(1 + q)(2 + 1368046q + 10903287q^2 + 1368046q^3 +
     2q^4)\cr
X_1 =&~ 1 + q\cr
X_2 =&~ -6(7 + 22q + 7q^2)\cr
X_3 =&~ 9(1 + q)(109 + 148q + 109q^2)\cr
X_4 =&~ -4(4247 + 28450q + 33606q^2 + 28450q^3 + 4247q^4)\cr
X_5 =&~ 3(1 + q)(81410 - 367682q - 3523185q^2 - 367682q^3 +
     81410q^4)\cr
Y_0 =&~ q/(1 + q)^2\cr
Y_1 =&~ -24(-1 + q)^2q/(1 + q)^3\cr
Y_2 =&~ -36(-1 + q)^2q(1 + 50q + q^2)/(1 + q)^4\cr
Y_3 =&~ -8(-1 + q)^2q
     (55 + 4890q + 23494q^2 + 4890q^3 + 55q^4)/(1 + q)^5\cr
Y_4 =&~ -6(-1 + q)^2q(2279 + 378364q + 2348113q^2 +
       5049976q^3 +\cr&~ 2348113q^4 + 378364q^5 + 2279q^6)/
   (1 + q)^6}}
\eqn\dumb{\eqalign{
case~c. & \cr
G_0 =&~ 1\cr
G_1 =&~ 8(1 + q)\cr
G_2 =&~ 8(3 + 28q + 3q^2)\cr
G_3 =&~ 32(1 + q)(1 + 186q + q^2)\cr
G_4 =&~ 8(3 + 20176q + 69500q^2 + 20176q^3 + 3q^4)\cr
G_5 =&~ 48(1 + q)(1 + 88890q + 641386q^2 + 88890q^3 + q^4)\cr
X_1 =&~ 1 + q\cr
X_2 =&~ -8(3 + q)(1 + 3q)\cr
X_3 =&~ 20(1 + q)(15 + 26q + 15q^2)\cr
X_4 =&~ -64(41 + 352q + 430q^2 + 352q^3 + 41q^4)\cr
X_5 =&~ 2(1 + q)(9063 - 82738q - 620882q^2 - 82738q^3 +
     9063q^4)\cr
Y_0 =&~ q/(1 + q)^2\cr
Y_1 =&~ -16(-1 + q)^2q/(1 + q)^3\cr
Y_2 =&~ (8(-1 + q)^2q(1 + 98q + q^2)/(1 + q)^4\cr
Y_3 =&~ -64(-1 + q)^2q
     (1 + 128q + 766q^2 + 128q^3 + q^4)/(1 + q)^5\cr
Y_4 =&~ -4(-1 + q)^2q(377 + 72740q + 471887q^2 +
       1126728q^3 + \cr&~471887q^4 + 72740q^5 + 377q^6)/
   (1 + q)^6}}
\eqn\dumb{\eqalign{
case~d. & \cr
G_0 =&~ 1\cr
G_1 =&~ 360q(1 + q)\cr
G_2 =&~ 1080q^2(211 + 872q + 211q^2)\cr
G_3 =&~ 720q^2(1 + q)(2565 + 828158q + 4510066q^2 +
     828158q^3 + 2565q^4)\cr
G_4 =&~ 360q^2(-673760 + 16098480q + 5996617977q^2 +
     52744187568q^3 + 103707768312q^4 + \cr&~ 52744187568q^5 +
     5996617977q^6 + 16098480q^7 - 673760q^8)\cr
X_1 =&~ (1 + q)^3\cr
X_2 =&~ 12(1 + q)^2(5 - 196q - 642q^2 - 196q^3 + 5q^4)\cr
X_3 =&~ 18(1 + q)(-85 - 6755q + 78932q^2 + 349843q^3 +
     682082q^4 + \cr&~349843q^5 + 78932q^6 - 6755q^7 - 85q^8)\cr
X_4 =&~ 16(17135 + 127035q + 4239030q^2 - 186863233q^3 -
     1965821823q^4 - \cr&~5845039290q^5 - 8195715660q^6 -
     5845039290q^7 - 1965821823q^8 - 186863233q^9 + \cr&~
     4239030q^10 + 127035q^11 + 17135)\cr
Y_0 =&~ q/(1 + q)^2\cr
Y_1 =&~ -60(-1 + q)^2q(1 + 10q + q^2)/(1 + q)^3\cr
Y_2 =&~ 90(-1 + q)^2q
     (57 + 274q - 7341q^2 - 22796q^3 - 7341q^4 + \cr&~
       274q^5 + 57q^6)/(1 + q)^4\cr
Y_3 =&~ 40(-1 + q)^2q(-16844 - 22047q + 1066354q^2 -
       38340920q^3 - 242849702q^4 - \cr&~ 428992850q^5 -
       242849702q^6 - 38340920q^7 + 1066354q^8 - 22047q^9 -
       16844q^10)/(1 + q)^5\cr
case~e. & \cr
G_0 =&~ 1\cr
G_1 =&~ 840q^3(1 + q)\cr
G_2 =&~ 840q^4(-2 + 28q + 1491q^2 + 3960q^3 + 1491q^4 +
     28q^5 - 2q^6)\cr
X_1 =&~ (1 + q)^7\cr
X_2 =&~ 2(1 + q)^6(3 - 46q + 434q^2 - 2562q^3 - 11466q^4 -
     2562q^5 + 434q^6 - 46q^7 + 3q^8)\cr
Y_0 =&~ q/(1 + q)^2\cr
Y_1 =&~ 2(-1 + q)^2q(-1 + 13q - 113q^2 - 638q^3 -
       113q^4 + 13q^5 - q^6)/(1 + q)^3 }}

\listrefs
\end